\documentclass{article}

\usepackage[utf8]{inputenc}
\usepackage{amsmath}
\usepackage{cite}
\usepackage{tikz}
\usetikzlibrary{tikzmark}

\usepackage{epsf,hyperref}
\usepackage{amssymb,ComplexSystems}
\usepackage{xcolor}

\newtheorem{thm}{Theorem}[section]
\newtheorem{defn}{Definition}[section]

\begin{document}

\title{Classification of non-Leibnizian Strings}

\author{\authname{Furkan Semih Dündar}\\[2pt] 
\authadd{Department of Mechanical Engineering, Amasya University}\\
\authadd{Amasya, Türkiye}\\[8pt]
\authadd{Wolfram Institute for Computational Foundations of Science}\\
\authadd{Illinois, USA}
}

\markboth{Complex Systems} 
{Sample Paper for Complex Systems} 

\maketitle

\begin{abstract}
  Leibnizian ideas in foundations of physics has enjoyed a renewed interest. Cyclic string ontology in Wolfram Model multiway rewriting systems has been found to encode finite dimensional quantum systems where what are called as Leibnizian strings have been employed. It has then become important to classify strings in terms of their ability to correspond to a monadic interpretation \`a la Leibniz. The definition of Leibnizian strings are rather quite non-linear and for that purpose we classify non-Leibnizian strings for which there is a simpler criterion on maximal neighborhoods of characters. If a string is not non-Leibnizian then it must be Leibnizian. We found out three canonical forms of non-Leibnizian strings and noted that two of them are context-free languages whereas one is not.
\end{abstract}


\section{Introduction}

Leibnizian ideas in foundations of physics, such as monads, have gained a revival in recent years, though somewhat limited. The origin of this trend can be traced back to Barbour and Smolin's work~\cite{barbour1992} and Barbour's Ref.~\cite{barbour2003deep}. There are studies \cite{dundar2024case,cs,rickles2023ruliology} that directly employ Leibnizian ideas and others \cite{smolin2021views,smolin2023views,smolin2016quantum} that considers the notion of a `view'. Here the term \emph{view} corresponds to the Leibnizian notion of the view of a monad.

In order to investigate Leibnizian ideas in physics, the cyclic string ontology is a useful ground to test ideas. It is not trivially simple yet it is not as complex as the hypergraph ontology of Wolfram Model~\cite{wm}. For a mathematical account of Wolfram Model see Refs.~\cite{arsiwalla2024pregeometric,arsiwalla2021homotopies}. Recently, a study~\cite{qstr_monads} has appeared that uses Leibnizian ideas in Wolfram Model multiway rewriting systems to encode finite dimensional quantum systems. In that study, only Leibnizian strings have been used. In essence in a Leibnizian string each character is distinguished locally by their neighborhoods (by their `views') whereas in \emph{non}-Leibnizian strings, there are at least two characters that have isomorphic neighborhoods at the maximum allowable radius. It has then become important to classify strings as either Leibnizian or non-Leibnizian and understand the nature of distinction between the two cases. In this paper we classify all non-Leibnizian strings over an alphabet $\Sigma$. We find out three possible forms of non-Leibnizian strings and  express that although two cases belong to context-free languages one case is not context-free and is rather context \emph{sensitive}.

Formal languages has an important place in discussion of pregeometry and representing fundamental physics \cite{formal}. In this respect, studies on Leibnizian or non-Leibnizian strings, which are formal languages, are important in order to fully investigate the indications of works such as Ref.~\cite{qstr_monads}.

The organization of the paper is as follows. In Section~\ref{sec:prelim} we provide preliminary background material. In Section~\ref{sec:classification} we classify all possible non-Leibnizian strings. In Section~\ref{sec:language} we consider the formal language of non-Leibnizian strings. In Section~\ref{sec:conclusion} we conclude the paper with directions for future studies.

\section{Preliminaries}\label{sec:prelim}

In this Section we provide preliminary content, mainly from Ref.~\cite{barbour1992}. We consider a set of strings over some alphabet $\Sigma$. Moreover we suppose that the strings are \emph{cyclic}. Meaning that the character after the last character is the first character and the character before the first character is the last character.

\begin{defn}[Maximum radius]
  For a string of length $L$, the maximum radius of a character is defined to be $m$. If $L$ is odd, then $2m+1 = L$ or if $L$ is even, $2m+2=L$.
\end{defn}

\begin{defn}[String isomorphism]
Two strings $S_1,S_2$ are deemed as \emph{isomorphic} if they are the same or one is the mirror symmetric counterpart of the other.
\end{defn}

\begin{defn}[non-Leibnizian strings]
A string $S$ is non-Leibnizian if two characters at different locations  share isomorphic neighborhoods with the maximum radius $m$, or if the string includes a single occurrence of a character.
\end{defn}

\begin{example}
  These are examples for non-Leibnizian strings:

  \begin{itemize}
  \item \texttt{aaaaaa}
  \item \texttt{abbbbb}
  \item \texttt{ababab}
  \item \texttt{abccba}
  \end{itemize}
  
\end{example}

If there is only one element in $\Sigma$ the strings are trivially non-Leibnizian. From now on we consider an alphabet that has at least two letters. For completeness, we provide a definition of Leibnizian strings.

\begin{defn}[Leibnizian strings]
If a string $S$ is \emph{not} non-Leibnizian it is called as Leibnizian.
\end{defn}

In Leibnizian strings, each character is distinguished by local neighborhoods of radius less than or equal to $m$. That property allows one to model Leibniz's \emph{monads} by defining the `view' of monad as the smallest neighborhood that is not isomorphic to other neighborhoods. Then due to Leibniz's \emph{identity of indiscernibles}\footnote{This principle basically means that if all the properties of two entities are the same, then they are the same entity. For a detailed discussion please refer to Ref.\cite{sep-identity-indiscernible}.} principle each monad is found to be unique as required. Hence Leibnizian strings can model monads whereas non-Leibnizian strings \emph{cannot} model monads because in those there are at least two characters whose neighborhoods are not distinguished even at the maximum radius. That would mean `equal' monads, which are not allowed in Leibnizian philosophy.

Because in the present study we focus only on non-Leibnizian strings and their classification, we do not consider Leibnizian strings, their monadic interpretation and the associated Barbour-Smolin-Deutsch variety~\cite{barbour1992} (the term is coined in Ref.~\cite{dundar2024case} in the form of an acronym: `BSD Variety'). For more please see Refs.~\cite{barbour1992,qstr_monads,dundar2024case,cs}.
  
\section{Classification of non-Leibnizian Strings}\label{sec:classification}

In this Section, we classify all non-Leibnizian strings. If a string includes a single occurrence of a letter then the string is trivially non-Leibnizian. From now on we neglect those cases and suppose that any letter that is included in the string occurs more than once. As mentioned previously in a non-Leibnizian string there are two characters with \emph{isomorphic} neighborhoods at the maximum radius $m$. Due to nature of string isomorphism there are two cases:

\begin{itemize}
\item (Case-I) Neighborhoods are equal to each other.
\item (Case-II) One neighborhood is the mirror symmetric equivalent of the other.
\end{itemize}

On the other hand, we differentiate the cases of strings with odd or even length. This is because the maximum neighborhood of a character in a string of odd length includes all the string, however the maximum neighborhood of a character in a string of even length leaves out a single character. The main reason is that the number of characters in a maximum radius neighborhood is $2m+1$ where as the string length is $2m+1$ and $2m+2$ for odd and even $L$ respectively.

We suppose that the first character and the $j$'th character has isomorphic neighborhoods. Without any loss of generality we assume $2\leq j\leq m+1$. The cases for $j > m+1$ can be obtianed by cyclically rotating the string until the $j$'th character is placed in the first location of the string. We will need a theorem about string commutation, which we provide below.

As a short note on notation, we use lower-case Greek letters to denote single characters (\emph{e.g.} $\sigma,\alpha,\beta$) and upper-case Latin letters to indicate strings (\emph{e.g.} $A,B,C$) which may be empty as well. When we place letters or strings sequentially we suppose that they are concatenated. Moreover $S^T$ corresponds to \emph{transpose} of a string $S$, that is, the string $S$ in reverse order.

\begin{thm}\label{thm:ABBA}
  Let $A,B$ be two strings over the alphabet $\Sigma$. If $AB = BA$ holds, then either $A = C^a, B = C^b$ for some string $C$ or $A = \alpha^{|A|}, B = \alpha^{|B|}$ for some $\alpha \in \Sigma$ where $a, b \geq 0$.
\end{thm}

\begin{proof}
  Let $c$ be the greatest common divisor of $|A|$ and $|B|$. First we suppose $c=1$. Without any loss of generality we can suppose $|A| > |B|$. When we map $A \to BA$ we obtain the same equation, namely $AB = BA$. As long as $|A| > |B|$ we can use this map and obtain $A = B^{a_1} A_1$ where $|A| = a_1 |B| + |A_1|$. Note that the greatest common divisor of $|B|$ and $|A_1|$ is still one. In this case we have $|B| > |A_1|$ and apply the same mapping as before and obtain $B = A_1^{b_1} B_1$ where $|B| = b_1 |A_1| + |B_1|$. At the end, we will obtain an equation of the form $A_i B_j = B_j A_i$ where one of $A_i$ or $B_j$ will consists of a single character which will cause $A_i$ and $B_j$ to consists of a repeating single character. When one goes in the reverse direction the result $A = \alpha^{|A|}$ and $B = \alpha^{|B|}$ is obtained for some $\alpha \in \Sigma$.

  We now focus on the second part of the proof. In this case $c > 1$ holds. We can write $|A| = ac, |B| = bc$. Because $c$ is the greatest common divisor, it is seen that $a$ and $b$ are co-prime. We can write $A = A_1 A_2 \ldots A_a$ and $B = B_1 B_2 \ldots B_b$ where each of $A_i$ and $B_j$ have the length $c$. Because all of these substrings have the same length we can consider $A$ and $B$ to be strings over some generalized alphabet $\Sigma'$ which consists of $A_i$ and $B_j$, formally $\Sigma' = \bigcup_i A_i \cup \bigcup_j B_j$. When considered in this way, we see that the only solution is $A = C^a$ and $B = C^b$ for some string $C \in \Sigma'$ of length $c$.
\end{proof}

\begin{corollary}
  If strings $A,B$ have equal number of letters and $AB = BA$, then $A = B$.
\end{corollary}

\subsection{Case-I}

In this subsection we consider non-Leibnizian strings that are deemed as such using the neighborhood isomorphism in Case-I. To iterate, if $N_1$ and $N_2$ are two maximal neighborhoods then we need to have $N_1 = N_2$.

\subsubsection{Odd $L$}

Here we consider strings of the form $S = \sigma S_1 \sigma S_2$. We suppose that the two $\sigma$ characters at two different locations (index-1 and index-$j$) have equal maximal neighborhoods. Without loss of generality we consider $2\leq j \leq m+1$. Larger $j$ values can be obtained by cyclic rotations of the string.

In order to illustrate the neighborhood structure of each $\sigma$, we repeat the string $S$ three times and obtain $S^3$. This representation is useful to see the neighborhoods of characters considering the cyclic boundary condition.

\vspace{1em}

\begin{equation}
  S^3 = \sigma S_1 \sigma S_2 \tikzmarknode{s1}{\sigma} S_1 \tikzmarknode{s2}{\sigma} S_2\sigma S_1 \sigma S_2
\end{equation}

\begin{tikzpicture}[remember picture, overlay]
  \draw[black] ([yshift=-0.1cm]s1.south) -- ++(0,-0.3cm) coordinate (s1b);
  \draw[black] (s1b) -- ++(-1cm,0) -- ++(0.5cm,0) node[below] {$L_1$};
  \draw[black] (s1b) -- ++(1cm,0) -- ++(-0.5cm,0) node[below] {$R_1$};
  \draw[black] ([yshift=0.1cm]s2.north) -- ++(0,0.3cm) coordinate (s2t);
  \draw[black] (s2t) -- ++(-1cm,0) -- ++(0.5cm,0) node[above] {$L_2$};
  \draw[black] (s2t) -- ++(1cm,0) -- ++(-0.5cm,0) node[above] {$R_2$};
\end{tikzpicture}

We indicate each neighborhood as $N_1 = L_1 \sigma R_1$ and $N_2 = L_2 \sigma R_2$ where $L_i,R_i$ are strings of lenght $m$. Because $2 \leq j \leq m+1$, the string $R_1$ must be of the form $S_1\sigma D_1$ for some $D_1$ and similarly $L_2$ must be of the form $A_2 \sigma S_1$ for some $A_2$. Equating $N_1 = N_2$, and noting that for odd $L$ maximal neighborhoods cover the whole string we can write:

\begin{equation}
  S = \sigma S_1 \sigma D_1 A_2 \sigma S_1
\end{equation}

When we cyclically shift $S$ towards right by $|\sigma S_1| = j-1$ amount, we should obtain the same string, since we are considering Case-I symmetries. Hence we obtain (where we let $S_3 = D_1 A_2$):

\begin{equation}
  \sigma S_1 \sigma S_3 \sigma S_1 = \sigma S_1 \sigma S_1 \sigma S_3
\end{equation}

This equation is in the form of $AB = BA$ that we considered before, where now $A = \sigma S_1$ and $B = \sigma S_1 \sigma S_3$. Here we have $|B| > |A|$. The length of $S_1$ is $j-2$ and the length of $S_3$ is $2(m-j+1)$. Hence we obtain $|A| = j-1$ and $|B| = 2m-j+2.$ If these two numbers are co-prime, then the only solution is $S_1 = \sigma \cdots \sigma$ and $S_3 = \sigma \cdots \sigma$ where in the end

\begin{equation}
  S = \sigma^L
\end{equation}

Now we suppose that these numbers are not co-prime and hence there is a string $C$ with at least two characters such that $\sigma S_1 = C^a$ and $\sigma S_1 \sigma S_3 = C^b$ for $b > a$. We see that $\sigma S_3 = C^{b-a}$. It is also evident that $C = \sigma D$ for some $|D|\geq 1$. In this case we obtain (after changes of notation):

\begin{equation}
  S = (\sigma S_1)^a
\end{equation}

where $a = L/(1+|S_1|)$ for any string $S_1$ such that $a$ is an integer.

\subsubsection{Even $L$}

In the case of even $L$ there is a subtlety. It is that in this case the maximal neighborhoods do not cover the whole string and leaves out exactly one character. The structure of neighborhoods is the same as in odd $L$ case that we just discussed. But in this case the whole string is in the form of $S = N_1 \alpha \sim N_2 \beta$ for some characters $\alpha,\beta$ where $\alpha \in N_2$ and $\beta \in N_1$. Let us write down the maximal neighborhoods:

\begin{equation}
  N_1 = A_2 \sigma S_1 \sigma S_1 \sigma D_1 
\end{equation}

and considering that the whole string is $N_1\alpha$ for some $\alpha \in \Sigma$,

\begin{equation}
  S = A_2 \sigma S_1 \sigma S_1 \sigma D_1  \alpha
\end{equation}

When we rotate $N_1$ to right by $j-1$ steps we should obtain $N_2 \beta$. For that purpose we let $S_1\sigma D_1 = E_1 F_1$ where $|F_1| = |S_1| - 1 = j-2$. We now have the following equality:

\begin{equation}
  F_1 \alpha A_2 \sigma S_1 \sigma E_1 = A_2 \sigma S_1 \sigma E_1 F_1 \beta
\end{equation}

If $\alpha = \beta$, this equation is in the form of $AB = BA$. If the greatest common divisor of $|F_1 \alpha|$ and $|\alpha A_2 \sigma S_1 \sigma E_1|$ is greater than 1, for some string $G_1$ we have $F_1\alpha = (G_1 \alpha)^a$ and $\alpha A_2 \sigma S_1 \sigma E_1 = (G_1\alpha)^b$. The first character of $G_1$ is $\sigma$. In this case we obtain a string $S$ of the followig form:

\begin{equation}
  S = (\sigma S_1)^a
\end{equation}

where $a = L/(1+|S_1|)$ for any string $S_1$ such that $a > 1$ is an integer. On the other hand if $|F_1 \alpha|$ and $|\alpha A_2 \sigma S_1 \sigma E_1|$ are co-prime then only solution is $\alpha = \beta = \sigma$ and all the string that appear is $\sigma^a$ for some suitable $a$ for each string. The solution is a uniform string:

\begin{equation}
  S = \sigma^L
\end{equation}

So far we have considered the case $\alpha = \beta$ and we now suppose $\alpha \neq \beta$. Let us remember the equation we need to solve:

\begin{equation}
  F_1 \alpha A_2 \sigma S_1 \sigma E_1 = A_2 \sigma S_1 \sigma E_1 F_1 \beta
\end{equation}

We propose that this equation has no solution if $\alpha \neq \beta$. For simplicity let us define $F \equiv F_1$ and $A \equiv A_2 \sigma S_1 \sigma E_1$. Then the equation is:

\begin{equation}
  F \alpha A = A F \beta
\end{equation}

The idea is to use recursion. It is evident that $|A| > |F|+1$. We map $A \to (F\alpha)^a A$ for some $a \geq 1$ and we obtain $F\alpha A = A F \beta$. Now $A < |F\alpha|$. We now map $F\alpha = A^b F\alpha$ for some $b \geq 1$ and obtain $F\alpha A = A F \beta$. This process ends if $A = \varepsilon$ or $F = \varepsilon$, where `$\varepsilon$' denotes the empty string. If $A = \varepsilon$ then we have $F\alpha = F\beta$ and $\alpha = \beta$. If $F = \varepsilon$ then we have $\alpha A = A \beta$. We have a solution if $\alpha = \beta$. If $\alpha \neq \beta$ then we have no solution for $A$. Let us see this in the following way. Since $\alpha A = A \beta$ we map $A \to \alpha^i \beta^j$ for some integers $i,j$ and we obtain $\alpha = \beta$. In both of these cases we get $\alpha = \beta$ which is a contradiction, since we supposed that $\alpha \neq \beta$. Hence in this case there is no solution.

\subsection{Case-II}

In this subsection we consider non-Leibnizian strings that are deemed as such using the neighborhood isomorphism in Case-II. To iterate, if $N_1$ and $N_2$ are two maximal neighborhoods then we need to have $N_1 = N_2^T$.

\subsubsection{Odd $L$}

We use the same parameterization of the string as $S = \sigma S_1 \sigma S_2$. We have $N_1 = L_1 \sigma R_1$ and $N_2 = L_2 \sigma R_2$. Because $2 \leq j \leq m+1$, $R_1 = S_1\sigma D_1$ and $L_2 = A_2 \sigma S_1$ for some strings $A_2, D_1$. Hence $N_1 = L_1 \sigma S_1 \sigma D_1$ and $N_2 = A_2 \sigma S_1 \sigma R_2$. Since for Case-II we have $N_1 = N_2^T$ we obtain:

\begin{equation}
  L_1 \sigma S_1 \sigma D_1 = R_2^T \sigma S_1^T \sigma A_2^T.
\end{equation}

Here we see that $S_1 = S_1^T$ because $|D_1| = |A_2^T|$ and $|L_1| = |R_2^T|$. On the other hand we have $S \sim N_1 \sim N_2$, which yields:

\begin{equation}
  \sigma S_1 \sigma A_2^T L_1 = \sigma S_1 \sigma L_1^T A_2.
\end{equation}

Observe that $A_2^T L_1 = L_1^T A_2$ meaning that $S_2 = A_2^T L_1$ is symmetric. Hence the result of this part is as follows:

\begin{equation}
  S = \sigma S_1 \sigma S_2
\end{equation}

where $S_1 = S_1^T$ and $S_2^T = S_2$. Since $L$ is odd, we also require that $|S_1| + |S_2|$ is an odd number.

\subsubsection{Even $L$}

The setup is the same as previous subsection on Case-I for even $L$. We write $S = \sigma S_1 \sigma S_2$. The neighborhoods are then $N_1 = L_1 \sigma S_1 \sigma D_1$ and $N_2 = A_2 \sigma S_1 \sigma R_2$. In Case-II we have $N_1 = N_2^T$ and this results in:

\begin{equation}
  L_1 \sigma S_1 \sigma D_1 = R_2^T \sigma S_1^T \sigma A_2^T
\end{equation}

Since $|D_1| = |A_2^T|$ we need to have $S_1^T = S_1$, $L_1 = R_2^T$ and $D_1 = A_2^T$. For some $\alpha,\beta \in \Sigma$ we have $S = N_1\alpha$ and $S = N_2 \beta$ where cyclic boundary condition is considered. When we shift $N_1\alpha$ by $|L_1|$ we should obtain $S$ as it is previously expressed in the beginning of this part. We also shift $N_2 \beta$ by $|A_2|$ we again obtain $S$. These indicate the following equation holds

\begin{equation}
  \sigma S_1 \sigma A_2^T \alpha L_1 = \sigma S_1 \sigma L_1^T \beta A_2
\end{equation}

and we have

\begin{equation}
  A_2^T \alpha L_1 = L_1^T \beta A_2
\end{equation}

If $\alpha = \beta$ we obtain $S_2 = A_2^T \alpha L_1$, $S_2^T = S_2$ and the form of $S$ is then expressed as

\begin{equation}
  S = \sigma S_1 \sigma S_2
\end{equation}

where $S_1^T = S_1$, $S_2^T = S_2$ and $|S_1| + |S_2|$ is even. Now suppose $\alpha \neq \beta$. Let us define $A \equiv A_2^T$ and $B = L_1$. We have

\begin{equation}
  A\alpha B = B^T \beta A^T
\end{equation}

We map $A \to \gamma A, B \to B\gamma$ for some $\gamma \in \Sigma$. Then the equation becomes $A\alpha B = B^T \beta A^T$ which has the same form. Since $|B| > |A|$ in the end we will have $A \to \varepsilon$ and $\alpha B = B^T \beta$ where $|B| \geq 1$. If we let $B \to B \alpha$ we get $B\alpha = B^T \beta$ which implies $\alpha = \beta$ which is a contradiction since we assumed $\alpha \neq \beta$. Hence there is no solution if $\alpha \neq \beta$. 

\subsection{All Possible Forms of non-Leibnizian Strings}

To summarize the results of this Section, we express that if a string $S$ is non-Leibnizian then it belongs to one of the following cases, where $\sigma \in \Sigma$: 1) $S$ has a character that occurs only once. $S = \sigma S_1$ for all $S_1$ such that $\sigma \notin S_1$, 2) $S$ has symmetric substrings. $S = \sigma S_1 \sigma S_2$ where $S_1^T = S_1$ and $S_2^T = S_2$, 3) $S$ is periodic. $S = (\sigma S_1)^a$ for some integer $a > 1$. We can further generalize these and write

\begin{enumerate}
\item $S$ has a character that occurs only once. $S = \sigma S_1$ for all $S_1$ such that $\sigma \notin S_1.$
\item $S$ has symmetric substrings. $S = S_1 S_2$ where $S_1^T = S_1$ and $S_2^T = S_2$.
\item $S$ is periodic. $S = S_1^a$ for integer $a > 1$ and string $S_1$.
\end{enumerate}

These are the \emph{canonical} forms of non-Leibnizian strings. One is allowed to generate others by cyclic string rotations. 

\section{Language of non-Leibnizian Strings}\label{sec:language}

Let us define the following sets:

\begin{align}
  nL_1 &= \{\sigma S \mid \sigma \in \Sigma, S \in \Sigma^*, \sigma \notin S\}\\
  nL_2 &= \{ S_1 S_2 \mid S_1,S_2 \in \Sigma^*, S_1^T = S_1, S_2^T = S_2\}\\
  nL_3 &= \{ S^a \mid a > 1, S \in \Sigma^*\}
\end{align}

It is easy to show that $nL_1$ and $nL_2$ are context-free languages. However $nL_3$ is not. We construct context-free grammars $G_i$ for $nL_i$ for $i = 1,2$. In these we consider an alphabet with two letters $\Sigma = \{a,b\}$. These grammars can be generalized for alphabets with more than two letters.

\begin{figure}[h]
  \centering
  \includegraphics{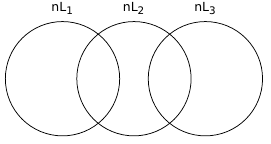}
  \caption{Venn diagram illustration of the sets.}
  \label{fig:venn}
\end{figure}

It is seen that these three sets are not mutually exclusive. In particular $nL_1 \cap nL_2 \neq \varnothing$, $nL_1 \cap nL_3 = \varnothing$ and $nL_2 \cap nL_3 \neq \varnothing$. For an illustration, please see Figure~\ref{fig:venn}.

In general $nL_1$ is not a proper subset of $nL_2$. However if there are only two letters in the alphabet, $nL_1 \subset nL_2$. Let $\Sigma = \{a,b\}$. The elements of $nL_1$ are of the form $a b\ldots b$ or $b a\ldots a$ which are always in $nL_2$ for $S_1 = a$ and $S_2 = b\ldots b$, or vice versa.

Let us now give examples that are at the intersection of two sets. Let $\sigma S \in nL_1$. If $S_1 = \sigma$ and $S_2 = S$ this string belongs to both of $nL_1$ and $nL_2$. If $S_1 = S_2 = S$ then the string $S_1 S_2$ is an element of both $nL_2$ and $nL_3$. There is no intersection between $nL_1$ and $nL_3$ because in elements of the former set a character occurs only once where as in the latter set the strings are periodic and for that purpose it does not include a single occurrence of a specific letter.

\subsection{The language $nL_1$}

$nL_1$ is a context-free language. The context-free grammar for the language $nL_1$ is as follows:

\begin{align}
  S &\to S_1 \mid S_2\\
  S_1 &\to a T_1\\
  S_2 &\to bT_2\\
  T_1 &\to b \mid T_1 T_1 \mid \varepsilon\\
  T_2 &\to a \mid T_2 T_2 \mid \varepsilon
\end{align}

\subsection{The language $nL_2$}

$nL_2$ is a context-free language. The context-free grammar for the language $nL_2$ is as follows:

\begin{align}
  S &\to S_1 S_2\\
  S_1 &\to a S_1 a \mid b S_1 b \mid \varepsilon\\
  S_2 &\to a S_2 a \mid b S_2 b \mid \varepsilon
\end{align}

\subsection{The language $nL_3$}

The language $nL_3$ is not a context-free language. We show that using what is called as \emph{the pumping lemma}. It is expressed as follows:

\begin{lemma}[The pumping lemma \cite{sipser}]
  Suppose that $L$ is a context-free language. Then there is $p > 0$ and for any string $S \in L$ with $|S| \geq p$ it can be written as $S = uvxyz$ such that the following conditions are satisfied:

  \begin{itemize}
  \item $\forall n \geq 0$, $uv^nxy^nz \in L$
  \item $|vy| > 0$
  \item $|vxy| \leq p$
  \end{itemize}
  
\end{lemma}

\begin{thm}
  The language $nL_3$ is \emph{not} context-free.
\end{thm}

\begin{proof}
  Let $p$ be the pumping length. We consider the string $S = a^p b^p a^p b^p$ and define $S_n \equiv u v^n x y^n z$. Initially we see that $vxy$ cannot be a substring of $a^p$ or $b^p$, otherwise $S_n \notin nL_3$. There are two cases: $vxy$ is a string of the form $a^i b^j$ or $b^i a^j$ for $2 \leq i+j \leq p$ with $i,j > 0$.

  We consider the first case, where $vxy$ is found in the first half of $S$ without any loss of generality. There are three cases:

    \begin{itemize}
    \item $v = a^i b^l, x = b^m, y = b^q$
    \item $v = a^k, x = a^l b^m, y = b^q$
    \item $v = a^k, x = a^l, y = a^m b^j$
    \end{itemize}

    When one considers $S_n$ with these parameters, one cannot obtain a periodic string. Hence $S_n \notin nL_3$. We now consider the last option: $vxy = b^i a^j$. We have three decompositions:

    \begin{itemize}
    \item $v = b^i a^k, x = a^m, y = a^q$
    \item $v = b^k, x = b^l a^m, y = a^q$
    \item $v = b^k, x = b^m, y = b^q a^j$
    \end{itemize}

    As well as in this case, we cannot obtian a periodic string $S_n$. Hence the string $S = a^p b^p a^p b^p$ cannot be pumped and the language $nL_3$ is not a context free language. The key idea here is that if $S_n \in nL_3$ then the string $S_n$ should begin with the letter `\texttt{a}' and end with the letter `\texttt{b}'. When one considers the form of decomposition in both cases, it is seen that $S_n = (a^p b^p)^{c_n}$ for some $c_n > 1$. For all the six decompositions this form is not obtained. Hence $nL_3$ is not a context-free language.

\end{proof}

\section{Conclusion}\label{sec:conclusion}

As Leibnizian ideas gain more visibility in foundations of physics it has become important to understand the distinction between Leibnizian and non-Leibnizian configurations, presently in the string ontology. It is a quite useful exercise of thought to consider the string ontology. It is not as complex as the (hyper-)graph ontology yet it is quite useful to be used to model physics.

Understanding the Leibnizian structure of strings is important in order to make contact with known physics. Especially see Ref.~\cite{qstr_monads} in this regard. That study considers Leibnizian strings to emulate finite dimensional quantum systems using Wolfram Model multiway rewriting systems.

In this study we concerned ourselves with the problem of classifying all non-Leibnizian strings. We have found three canonical cases for the form of such strings. We have determined that the first two are context-free languages whereas the third class could not be expressed by a context-free language.

As a future research direction, one may consider classifying all non-Leibnizian (hyper-)graphs. That will contribute to understanding the Leibnizian nature of various ontologies.

\bibliographystyle{unsrt}
\bibliography{refs}

\end{document}